\documentclass[aps,prd,twocolumn,superscriptaddress,nofootinbib]{revtex4-1}

\pdfoutput=1

\usepackage{amsmath}
\usepackage{amsfonts}
\usepackage{amssymb}
\usepackage[british]{babel}
\usepackage{graphicx}
\usepackage[T1]{fontenc}
\usepackage[utf8]{inputenc}
\usepackage[unicode=true,pdfusetitle,
bookmarks=true,bookmarksnumbered=true,bookmarksopen=true,bookmarksopenlevel=1,
breaklinks=false,pdfborder={0 0 0},backref=false,colorlinks=true]
{hyperref}
\hypersetup{
  citecolor=blue,filecolor=blue,linkcolor=blue,urlcolor=blue}
\usepackage{color}
\usepackage{ulem}

\definecolor{darkgreen}{rgb}{0,0.6,0}

\def\be{\begin{equation}}
  \def\ee{\end{equation}}
\def\bea{\begin{eqnarray}}
  \def\eea{\end{eqnarray}}
\def\ba{\begin{align}}
  
\def\bi{\begin{itemize}}
  \def\ei{\end{itemize}}

\def \si {\sigma}

\def \De {\Delta}

\def \om {\omega}
\def \Om {\Omega}

\newcommand{\MM}{{\mathcal M}}

\begin{document}


\title{Lensing convergence and the neutrino mass scale
in galaxy redshift surveys}


\author{Wilmar Cardona}
\affiliation{D\'epartement de Physique Th\'eorique and Center for Astroparticle Physics, Universit\'e de Gen\`eve, 24 quai Ernest Ansermet, CH--1211 Gen\`eve 4, Switzerland}

\author{Ruth Durrer}
\affiliation{D\'epartement de Physique Th\'eorique and Center for Astroparticle Physics, Universit\'e de Gen\`eve, 24 quai Ernest Ansermet, CH--1211 Gen\`eve 4, Switzerland}

\author{Martin Kunz}
\affiliation{D\'epartement de Physique Th\'eorique and Center for Astroparticle Physics, Universit\'e de Gen\`eve, 24 quai Ernest Ansermet, CH--1211 Gen\`eve 4, Switzerland}

\author{Francesco Montanari}
\affiliation{Helsinki Institute of Physics and Physics Department, University of Helsinki\\
  P.O. Box 64, FIN-00014, University of Helsinki, Finland}
\affiliation{D\'epartement de Physique Th\'eorique and Center for Astroparticle Physics, Universit\'e de Gen\`eve, 24 quai Ernest Ansermet, CH--1211 Gen\`eve 4, Switzerland}

\date{\today}

\begin{abstract}
  We demonstrate the importance of including the lensing contribution in galaxy clustering analyses with large galaxy redshift surveys. It is well known that radial cross-correlations between different redshift bins of galaxy surveys are dominated by lensing. But we show here that also neglecting lensing in the autocorrelations within one bin severely biases cosmological parameter estimation with redshift surveys. It leads to significant shifts for several cosmological parameters, most notably  the scalar spectral index and the neutrino mass scale. Especially the latter parameter is one of the main targets of future galaxy surveys.
\end{abstract}


\maketitle

\section{Introduction}

Galaxy number counts are a key observable in cosmology and are exploited by current \cite{Alam:2015mbd,Drinkwater:2009sd,Abbott:2005bi} and future
\cite{Dawson:2015wdb,Laureijs:2011gra,Abell:2009aa,Maartens:2015mra}
cosmological observations. Usually, the galaxy counts are compared to the predicted power spectrum
of matter density fluctuations $P(k,z)$, which is, however, not  directly observable. The power spectrum in harmonic space
$C_\ell(z,z')$ on the other hand is an observable \cite{Bonvin:2011bg}.

Predictions for  number counts have been derived for example in Refs. \cite{Yoo:2009au,Bonvin:2011bg,Challinor:2011bk}. From these expressions, it becomes clear that, in addition to the well-known density perturbations and redshift space distortions that are usually included in $P(k)$, there are additional, so-called relativistic effects that contribute to the observed number counts. They come from the fact that we observe photons which have been deflected on their way from a source into the telescopes and that not only the galaxy number but also the volume is perturbed.
For typical surveys, the most important relativistic effect is due to lensing convergence \cite{Bartelmann:1999yn,Matsubara:2000pr}.
In this paper, we show that neglecting lensing convergence in the analysis of a future survey like Euclid will lead to significant biases in the estimation of cosmological parameters. As a consequence, care should be taken to include lensing. Including it in the standard matter power spectrum $P(k)$ is difficult as lensing inherently mixes different scales; this
argues in favor of the adoption of quantities like the $C_\ell(z,z')$ where it is straightforward to include relativistic effects.

\section{Methodology}

We illustrate the bias of cosmological parameters when neglecting lensing by analyzing $C_\ell(z,z')$.
We employ CLASSgal \cite{DiDio:2013bqa} to compute ``observed'' $C_\ell$ that include the effect of lensing convergence in addition to the
matter perturbations and to redshift space distortions and ``theory'' $C_\ell$ that only contain the latter two and
neglect  lensing. We call the former $C_\ell^{\rm obs}$ and the latter $C_\ell^{\rm th}$.
In order to mimic a $P(k)$ analysis more closely, we consider not only the full set of $C_\ell(z,z')$ but add
a case where we limit ourselves to the autocorrelations $C_\ell(z,z)$.
More details about power spectra are given in Appendix~\ref{apa}.

The survey configuration which we consider here is consistent with the Euclid photometric catalog.
The number of galaxies per redshift and per steradian, the galaxy density, and the magnification bias are as specified in Ref.~\cite{Montanari:2015rga}. In order to make our work more self-contained,  we repeat them in Appendix~\ref{apa}.  For the galaxy bias, we assume $b_{\rm G}=b_0\sqrt{1+z}$ \cite{Amendola:2016saw}, where $b_0$ is varied in the Markov chain Monte Carlo (MCMC) chains.

We adopt a covered sky fraction $f_{\rm sky}=0.364$ and divide the $N\sim10^9$ photometric galaxies catalog into $N_{\rm bin}=5$ Gaussian redshift bins containing equal numbers of galaxies per steradian ${\cal N}$. We assume a fiducial flat $\Lambda$CDM model consistent with Planck \cite{Ade:2015xua}, including massive neutrinos with a normal mass hierarchy (dominated by the heaviest neutrino mass eigenstate).  More precisely, the cosmological parameters of our fiducial model are the reduced baryon density parameter,
$h^2\Om_b=\omega_{b } = 2.225 \times 10^{-2}$; the cold dark matter density parameter,
$h^2\Om_{\rm cdm }=\omega_{\rm cdm } = 0.1198$; the scalar spectral index,
$n_s = 0.9645$; the amplitude of curvature fluctuations,
$\ln10^{10}A_s = 3.094$; the Hubble constant
$H_0 = 67.27$km/s/Mpc $=h100$km/s/Mpc; and the sum of the neutrino masses,
$\sum m_{\nu} = 0.06$ eV and
$b_0=1$.

We incorporate an error $E_\ell^{ij}$ due to nonlinearities, computed as a rescaling of the transfer functions based on the {\sc Halofit} corrections to the power spectrum
(see Appendix D of Ref. \cite{Montanari:2015rga} for details; we neglect the parameter dependence of the error $E_\ell^{ij}$).
We also add a shot-noise contribution $\mathcal{N}^{-1}$ to the power spectra. Thus, the angular power spectrum of number count fluctuations is modelled as
\begin{equation} \label{eq:Cl_th_obs}
  C_\ell^{\mathrm{A},ij} = C_\ell^{ij} + E_\ell^{ij} + \mathcal{N}^{-1} \delta_{ij},
\end{equation}
where $\mathrm{A=obs,th}$ and $i,j=1,...,N_{\rm bin}$ are redshift bin indices.

Similarly to the cosmic shear implementation of Ref. \cite{Audren:2012vy}, we adopt a Gaussian likelihood which leads to a $\chi^2$ relative to the fiducial model given by
\begin{equation} \label{eq:chi2}
  \Delta \chi^2 = \sum_{\ell=2}^{\ell_{\max}} (2\ell+1) f_{\rm sky} \left( \ln \frac{d_\ell^{\rm th}}{d_\ell^{\rm obs}} + \frac{d_\ell^{\rm mix}}{d_\ell^{\rm th}} - N_{\rm bin}\right),
\end{equation}
where $d_\ell^{\mathrm{A}} \equiv \det ( C_\ell^{\mathrm{A},ij} )$ and $d_\ell^{\rm mix}$ is computed like $d_\ell^{\rm th}$ but substituting in each term of the determinant  one factor by $C_\ell^{\mathrm{obs},ij}$.  The total  $d_\ell^{\rm mix}$ is obtained by adding all different possibilities for the insertion of $C_\ell^{\mathrm{obs},ij}$. More details can be found in Ref.\cite{Audren:2012vy}. To be conservative and keep nonlinear effects small, we choose $\ell_{\rm max}=400$ in the analysis.

Angular power spectra $C_\ell$ and nonlinear corrections $E_\ell$  are accurately computed using the Limber approximation only for the lensing integral along the line of sight. We then explore the parameter space with the help of a MCMC approach based on the Metropolis-Hastings algorithm~\cite{Lewis:2002ah} first using wide flat priors (``without priors'') and a second time using Planck \cite{Ade:2015xua} priors (``Planck priors''). When computing the theoretical spectra $C_\ell^{\rm th}$ with which we want to fit the observed $C_\ell$, we neglect lensing convergence. Our aim is to test the shift (bias) of cosmological parameters due to this mistake.
To speed up the MCMC exploration of parameter space, the $C_\ell$  of the theoretical spectra are computed less accurately than $C_\ell^{\rm obs}$, but we request that $\Delta \chi^2 \lesssim 0.2$ for the fiducial parameters when lensing is included in the analysis. Hence, the inaccuracy in our calculations can lead to an uncertainty of the order of $\Delta \chi^2 \lesssim 0.2$.

\section{Results}
In this section, we present the results of our analysis. We first study the case (nearly) without prior knowledge and compare the results with a Fisher matrix based analysis. Then, we introduce Planck priors, and in a final subsection we analyze what our results mean for the significance of the detection of the lensing term in the Euclid photometric survey.

\subsection{MCMC without priors}
We first determine the bias of the parameters due to neglecting the lensing term assuming nearly no prior knowledge. Of course, we have to assume some priors for the MCMC chain, but they are very wide and flat.
\begin{table}[!t]
  \centering
  \begin{tabular}{@{}cccccc}
    \hline
    \multicolumn{6}{c}{i) Consistently including lensing: $\Delta \chi^2 = 0$} \\
    \hline
    Parameter & Mean & Best fit & $\sigma$ &\hspace{-0.52cm} shift: Mean & Best fit \\
    \hline
    $\omega_b$ & $0.02979$ & $0.02285 $ &$0.00624 $ &  \quad$1.2\sigma$ & $ 0.1\sigma$ \\
    $\omega_{cdm}$ & $0.1455 $ & $0.1219 $ & \quad$0.0200 $ &  \quad$1.3\sigma$ & $0.1\sigma$ \\
    $n_s$      & $0.9476 $ & $0.9642 $ & $0.0387 $ &  \quad$0.4\sigma$ & $ <0.1\sigma$ \\
    $\ln10^{10}A_s$ & $3.047 $ & $3.097$ & $0.065 $ &  \quad$0.7\sigma$ & $ <0.1\sigma$ \\
    $H_0\left(\frac{\text{km}}{\text{s}\cdot\text{Mpc}}\right)$      & $73.84$ & $67.84$ & $5.48$ &  \quad$1.2\sigma$ & $ 0.1\sigma$ \\
    $m_{\nu}$\,(eV)  & $0.29$ & $0.09$ & $0.19$ & \quad $ 1.2\sigma$ & $ 0.2\sigma$ \\
    $b_0$ & $1.018$ & $1.000$ & $0.031$ & \quad$0.6\sigma$ & $<0.1\sigma$ \\
  \end{tabular}
  \begin{tabular}{@{}cccccc}
    \hline
    \multicolumn{6}{c}{ii) Neglecting lensing: $\Delta \chi^2 = 2064$} \\
    \hline
    Parameter & Mean & Best fit & $\sigma$ & \hspace{-0.52cm} shift: Mean & Best fit \\
    \hline
    $\omega_b$ & $0.02494$ & $0.02120 $ & $0.00556 $ &  \quad$0.5\sigma$ & $0.1\sigma$ \\
    $\omega_{cdm}$ & $0.1532$ & $0.1435$ & $0.0208$ &  \quad$1.6\sigma$ & $1.1\sigma$ \\
    $n_s$      & $0.8702$ & $0.8837$ & $0.0446$ &  \quad$2.1\sigma$ & $1.8\sigma$ \\
    $\ln10^{10}A_s$ & $ 2.867 $ & $2.965 $ & $ 0.394 $ &  \quad$0.6\sigma$ & $0.3\sigma$ \\
    $H_0\left(\frac{\text{km}}{\text{s}\cdot\text{Mpc}}\right)$      & $68.73$ & $66.76$ & $5.14$ &  \quad$0.3\sigma$ & $0.1\sigma$ \\
    $m_{\nu}$\,(eV)  & $0.43$ & $0.41$ & $0.16$ &  \quad$2.3\sigma$ & $2.2\sigma$ \\
    $b_0$ & $1.293$ & $1.200$ & $0.271$ & \quad$1.1\sigma$ & $0.7\sigma$\\
  \end{tabular}
  \begin{tabular}{@{}cccccc}
    \hline
    \multicolumn{6}{c}{\parbox[t]{4.4cm}{iii) Neglecting lensing: \\ \hspace*{0.9cm} (only autocorrelations)} $\Delta \chi^2 = 180$} \\
    \hline
    Parameter & Mean & Best fit & $\sigma$ & \hspace{-0.52cm} shift: Mean & Best fit\\
    \hline
    $\omega_b$ & $0.01982 $ & $0.01737 $ & $0.00520 $ &  \quad$0.5\sigma$ & $0.9\sigma$ \\
    $\omega_{cdm}$ & $0.1658 $ & $0.1552 $ & $0.0242 $ &  \quad$1.9\sigma$ & $1.5\sigma$ \\
    $n_s$      & $0.7539 $ & $0.7675 $ & $0.0513 $ &  \quad$4.1\sigma$ & $3.8\sigma$ \\
    $\ln10^{10}A_s$ & $2.449 $ & $2.719 $ & $0.465 $ &  \quad$1.4 \sigma$ & $0.8\sigma$ \\
    $H_0\left(\frac{\text{km}}{\text{s}\cdot\text{Mpc}}\right)$      & $61.64 $ & $59.11$ & $5.43$ &  \quad$1 \sigma$ & $1.5\sigma$ \\
    $m_{\nu}$\,(eV)  & $0.41$ & $0.41$ & $0.14$ &  \quad$2.6\sigma$ & $2.5\sigma$ \\
    $b_0$ & $1.888$ & $1.603$ & $0.428$ & \quad$2.1\sigma$ & $1.4\sigma$ \\
  \end{tabular}

  \caption{MCMC results (flat prior). We show the mean and best-fit values, the standard deviation, and the amplitude of the shift of the mean and best fit with respect to the fiducial value in units of the standard deviation, $\si$, of the corresponding analysis. The large value of $\De\chi^2$ for case ii shows that cross-correlations cannot be fitted if lensing is neglected.  A shift of less than about $0.2\si$ is not serious and is probably due to the reduced precision used to compute the theoretical spectra.
  }
  \label{Table:1}
\end{table}

\begin{figure*}[bthp]
  \centering
  \includegraphics[scale=1.2]{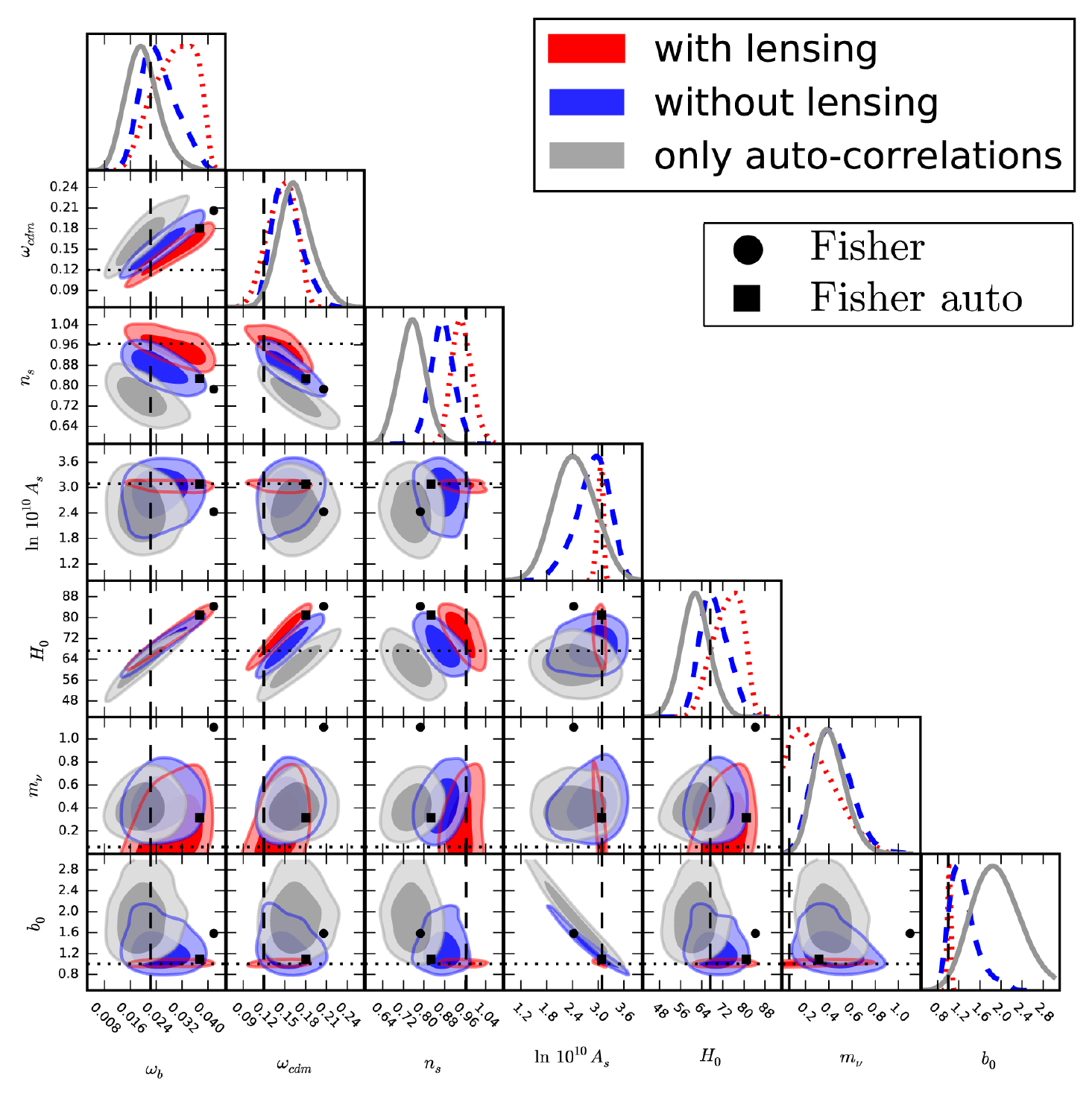}
  \caption{Two- and 1-D posteriors for the cosmological parameters inferred from the full analysis including lensing (red dotted), an analysis neglecting lensing (blue dashed) and considering only autocorrelations (gray solid).
    The 68\% and 95\% confidence intervals are shown.
    Intersections between vertical and horizontal lines denote the fiducial cosmology.
   In this analysis no significant priors were imposed on the parameters.  Circles and squares represent the estimates for the best fits from a Fisher matrix analysis when neglecting lensing, and for the only autocorrelations case, respectively.
  }
  \label{fig:mcmc}
\end{figure*}

We have  fitted the generated $C_\ell^{ij}$ data in three different ways,  where lensing convergence is i) consistently included, ii) neglected, and iii) neglected with only redshift bin autocorrelations are taken into account. The results are shown in Table~\ref{Table:1} and Fig.~\ref{fig:mcmc}. Figure \ref{fig:mcmc} shows two-dimensional contours and one-dimensional (1D) probability distribution functions for the marginalized posteriors of the cosmological parameters obtained from these analyses. The red contours (dotted 1D distributions) show the full analysis. They should reproduce the fiducial model. In the analyses shown by the gray (1D solid) and blue (1D dashed) contours, lensing is neglected. Furthermore, in the gray  contours, only autocorrelations [i.e.,\ $C_\ell(z,z)$] are considered, while the blue contours use both auto and cross-correlations [i.e.,\ $C_\ell(z,z')$ for all combinations of redshift bins].
The autocorrelation case is closer to the standard $P(k)$ analysis which is usually performed in redshift bins, but caution should be taken in comparing the two analyses since binning in redshift has significantly different effects.

From the red contours in Fig.~\ref{fig:mcmc}, it is evident that we cannot determine the baryon and cold dark matter densities very well with our configuration. The rather large redshift bins of our analysis with $\De z\gtrsim 0.3$ significantly smear out the baryon acoustic oscillations, leaving only the dominant features in the power spectrum which are fixed by the equality scale $k_{\rm eq} \propto \om_m/H_0$ (at fixed radiation content and measured in $h/$Mpc) and the ratio $\om_b/\om_{cdm}$. This  leads to a significant  degeneracy between $\om_b$, $\om_{\rm cdm}$, and $H_0$; only the slopes of the $(\om_x,H_0)$ and the $(\om_b, \om_{\rm cdm})$  contours are well determined. The large uncertainties in these parameters, as well as the prior $m_\nu\geq 0$, push the posterior mean value away from the best fit (which is always very close to the input value). We did not add realization noise in our likelihood, since it is not relevant for the present study.
Our aim here is not to derive optimal parameter constraints but to demonstrate the importance of the lensing contribution in such an analysis. For this reason, our approach is far from optimal but  conservative and simple, and even in this case, we find that not including lensing leads to wrong results.
Optimizing error contours by, e.g., introducing more nonlinear scales in the analysis is expected to lead to even more biased results, given  that  the relevance of lensing increases at higher multipoles.

If lensing is neglected in the analysis, several parameters
show a significant bias with respect to the input parameters (given by the vertical dashed lines);  cf. also Table~\ref{Table:1}.
First of all, there is a very strong degeneracy between the scalar amplitude $A_s$ and the bias $b_0$.  When including lensing which does not depend on $b_0$, this degeneracy is broken, and both $b_0$ and $A_s$ are determined accurately. Furthermore, lensing (together with the magnification bias for Euclid specifications) enhances clustering. Compensating this with a larger value of $b_0^2A_s$ leads to too much clustering on small scales, which, in turn, is compensated by reducing the spectral index by $(2-4)\sigma$ and by increasing the neutrino mass.   The preferred neutrino mass
is around $0.4\, \mathrm{eV}$, which corresponds just about to the current limits from cosmology~\cite{Ade:2015xua}.
From the degeneracy directions in the two-dimensional contours in Fig.~\ref{fig:mcmc},
we can also read off that forcing $m_\nu \rightarrow m_\nu^{\rm (fid)} = 0.06\, \mathrm{eV}$ would lead to
an even larger bias in the scalar spectral index.
Therefore, to go beyond the current state and derive accurate estimates of the neutrino masses with galaxy surveys absolutely requires taking lensing into account.

For models with additional parameters, the possibilities to improve the fit by choosing ``wrong'' values of the parameters increase, and we may see even larger biases, leading to even stronger spurious detections of new physics.

\begin{figure}[tp]
  \centering
  \includegraphics[width=\columnwidth]{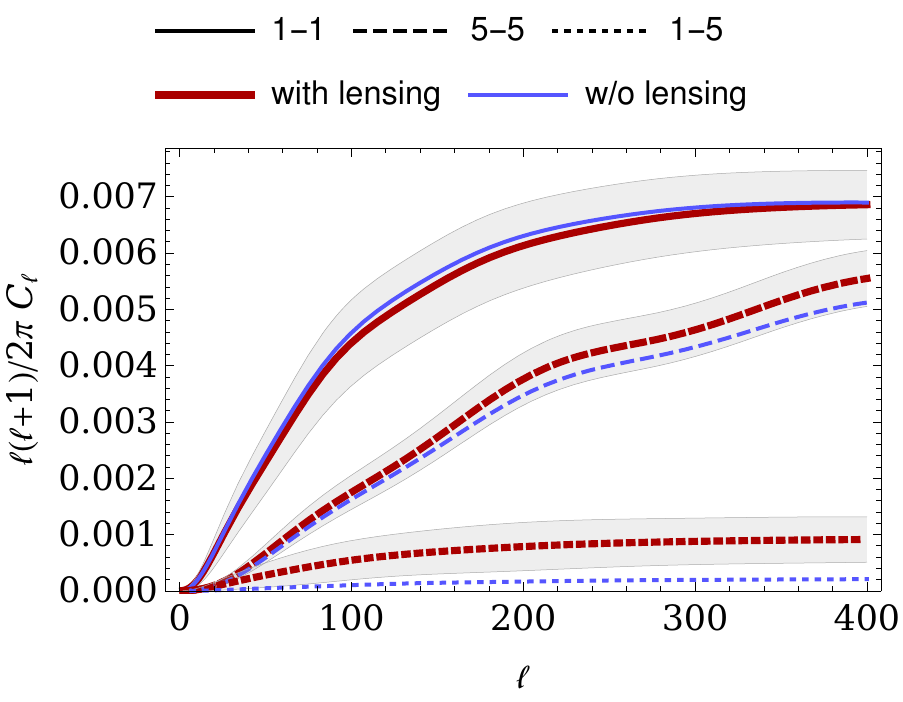}
  \caption{
    The thick red and the thin blue lines correspond to the spectra at the best-fit values estimated by consistently including lensing and by neglecting it, respectively.
    Gaussian error bars accounting for cosmic variance and shot noise for the consistent analysis are shown as gray regions.
    The indices for the correlated redshift bins are shown in the legend.
    The model neglecting lensing cannot fit the data, especially due to redshift cross-correlations.
  }
  \label{fig:cl}
\end{figure}

It is interesting that taking cross-correlations into account helps somewhat to reduce the bias on the parameters. The scalar spectral index best fit in this case has a bias of $1.8\sigma$,  as compared to $3.8\sigma$ for autocorrelations only, and the neutrino mass is shifted by $2.2 \sigma$ compared to $2.5\sigma$;
see Table~\ref{Table:1}. But this ``improvement'' is actually not real. It comes to a big extent from the fact that cross-correlations simply cannot be fitted without the lensing term as discussed below and shown in Fig.\ \ref{fig:cl}.  This is most manifest in the  total $\De\chi^2$ which increases from $\De\chi_{\rm auto}^2\simeq 180$ for the five autocorrelation bins to more than $\De\chi_{\rm a+c}^2\gtrsim2000$ when adding the ten cross-correlation bins. Giving each bin naively the same weight, we would expect an increase by a factor 3; instead, we have
$\De\chi_{\rm a+c}^2/\De\chi_{\rm auto}^2\gtrsim 11$.
The increase in the size of the parameter contours for some parameters appears at first counterintuitive as including more data improves our knowledge and therefore should reduce the errors.
This simple logic, however, only applies if the data can actually be fitted by the model at hand or if the likelihoods are Gaussian. Otherwise, different data may prefer different model parameters and lead to an increase not only in the total $\De\chi^2$ but also in the size of the confidence contours.

We can understand our results by looking at the differences in the harmonic power spectra shown in Fig.\ \ref{fig:cl}.
The thick red and thin blue lines are the angular spectra computed at the best-fit values shown in Table~\ref{Table:1} for the consistent case including lensing and for the one neglecting it, respectively (we include all redshift bin correlations).
For the consistent spectra including lensing, we compute 1-$\sigma$ error bars at each multipole by assuming, as for Eq.~(\ref{eq:chi2}), Gaussian spectra (see Eq.~(2.13) of Ref. \cite{Montanari:2015rga}).
We consider the representative correlations between the redshift bins $(ij)=(11), ~(55)$, and $(15)$ .
The plot shows that, when neglecting lensing, the spectrum for the cross-correlation between redshift bins 1 and 5 lies outside the 1-$\sigma$ error bars around the fiducial spectrum including lensing. This confirms that the model cannot fit the mock data.

\begin{table}[!t]
  \centering
  \begin{tabular}{c|cccc}
    Parameter & \multicolumn{2}{c}{Shift of best fit } &\multicolumn{2}{c}{for MCMC} \\
    \hline
       $\omega_b$ & $1.2\sigma$ & ($0.9\sigma$) & $-0.1\si$ & ($-0.9\si$)\\
   $\omega_{cdm}$ & $1.7\sigma$ & ($1.1\sigma$)& $1.1\sigma$ &
($1.5\sigma$)\\
   $n_s$      & $-1.9\sigma$ & ($-1.3\sigma$) & $-1.8\sigma$ &
($-3.8\sigma$) \\
   $\ln10^{10}A_s$ & $-1.1\sigma$ & ($0.005\sigma$)& $-0.3\sigma$ &
($-0.8\sigma$)\\
   $H_0\left(\frac{\text{km}}{\text{s}\cdot\text{Mpc}}\right)$ &
$1.2\sigma$ & ($0.9\sigma$)& $-0.1\sigma$ & ($-1.5\sigma$)\\
   $m_{\nu}$\,(eV)  & $3.3\sigma$ & ($0.6\sigma$)  & $2.2\sigma$ &
($2.5\sigma$)\\
   $b_0$      & $1.7\sigma$ & ($0.1\sigma$)   & $0.7\sigma$ &
($1.4\sigma$)\\
  \end{tabular}

  \caption{
      Fisher matrix results for the shift in the best-fit values due to neglecting lensing, in units of standard deviations (see Fig. \ref{fig:fisher}). The numbers in parentheses refer the the case including only bin autocorrelations. For comparison we also give in columns 4 and 5 the corresponding values from the MCMC analysis presented in Table~\ref{Table:1} and Fig.~\ref{fig:mcmc}. While Fisher matrices give a good qualitative description of parameter degeneracies, estimates of the shifts in the best fits seriously misestimate the magnitude and direction in parameter space.
 }
  \label{Table:Fisher}
\end{table}

\begin{figure*}[bthp]
  \centering
  \includegraphics[width=\textwidth]{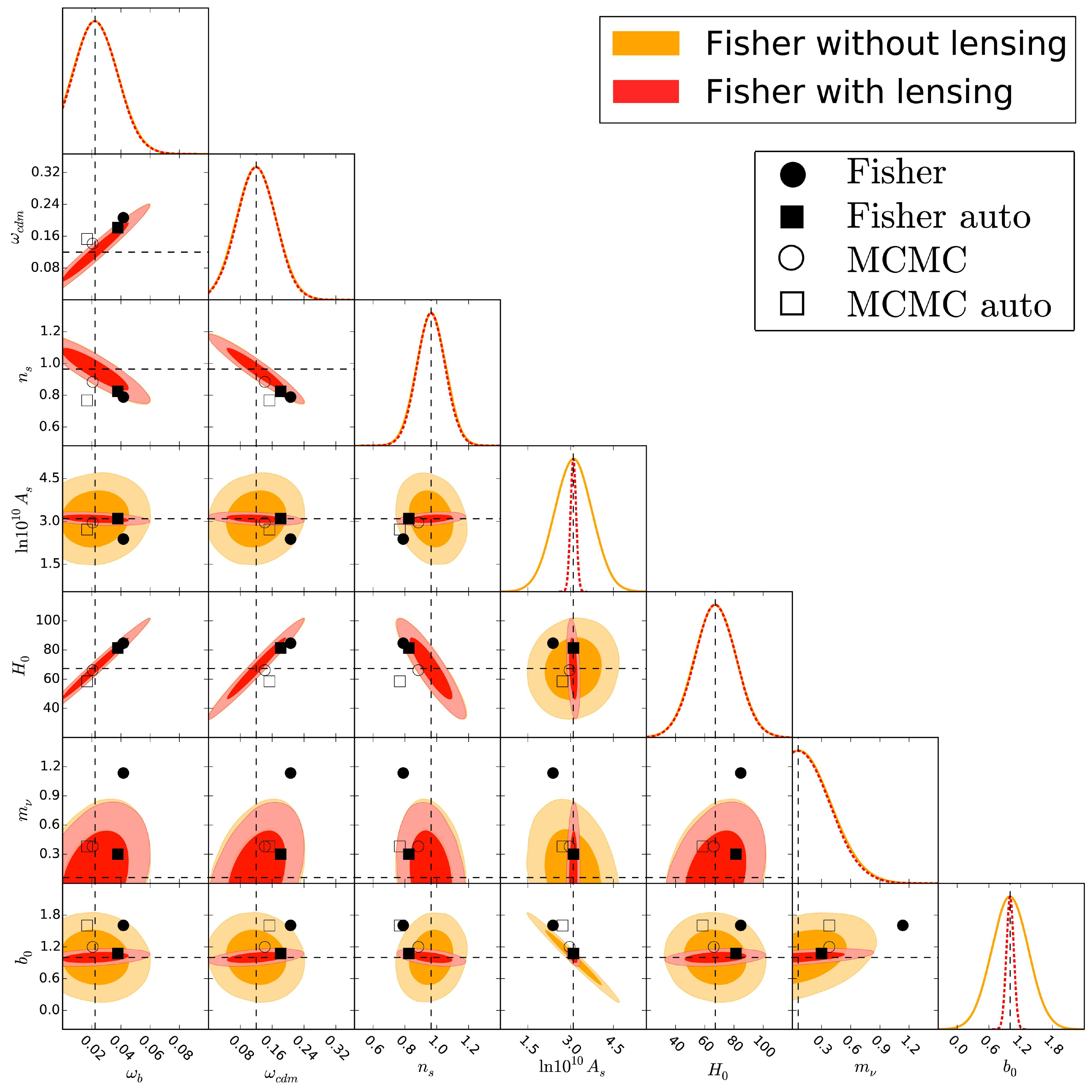}
  \caption{
      2D and 1D posteriors for the cosmological parameters inferred from the Fisher analysis excluding (orange solid) and including (red dotted) lensing.
    We stress that, in the former case, to compute error ellipses within the Fisher formalism, we forecast parameter constraints in a universe where lensing is absent (see the text for more details).
    The 68\% and 95\% confidence intervals are shown.
    Intersections of dashed lines denote the fiducial cosmology.
The expected systematic shifts in the best fit due to neglecting lensing in the theoretical modeling are shown, including all bin correlations (circles) and including only autocorrelations (squares).
For comparison, we also show the corresponding results from the MCMC analysis.
While the Fisher formalism is reliable for a qualitative understanding of parameter degeneracies, the systematic errors are seriously misestimated.
    See Table \ref{Table:Fisher} for more details about statistical quantities.
}
  \label{fig:fisher}
\end{figure*}

\subsection{Fisher analysis without priors}\label{s:fisher}

The shift of best-fit parameters  and the change in the figure of merit due to neglecting relativistic corrections (hence, in particular, neglecting lensing) has been studied previously; see, e.g.,  Refs. \cite{Namikawa:2011yr,Duncan:2013haa,Camera:2014sba,Raccanelli:2015vla}. However, these previous works did not include massive neutrinos, and they used a Fisher matrix analysis which gives quantitative estimates for shifts only if these are significantly less than one standard deviation.
Hence, the results obtained in these works can only be trusted qualitatively, while the MCMC study presented here gives quantitative results and demonstrates that large biases, exceeding by far $1\sigma$, are to be expected even for not very ambitious survey specifications.

We illustrate this here by repeating our analysis with a Fisher matrix technique.
The results are presented in Fig. \ref{fig:fisher}, where we also show the shifts of the best-fit values estimated via Fisher matrices (see Appendix~\ref{app:fisher} for details).
  Fisher matrix contours are reported for both cases, without and including lensing.
  This information is also reported in Table \ref{Table:Fisher}, where the standard deviations, $\si$, refer to the case without lensing, which provides more conservative information about the importance of the systematic error.
  We stress that in both cases, without and with lensing, Fisher matrices only forecast error contours around a universe described by the fiducial parameters and assume a Gaussian likelihood.
  While the MCMC analysis allows us to fit the wrong or the correct model to the data, in the Fisher context, this is not possible.
  This means that in the Fisher formalism we predict the error contours for both a model and a universe without lensing.
  Hence, while the red dotted contours with lensing can be compared between Figs. \ref{fig:mcmc} and \ref{fig:fisher}, the other contours have no correspondence between the two figures.
  In our case, Fisher matrices provide a good qualitative description of degeneracy between different parameter constraints. The 68\% confidence intervals are in disagreement with MCMC results by a factor 2--3, but the shapes and inclinations of the ellipses very roughly follow the MCMC contours.
  However, the magnitude and direction of the best-fit shift in parameter space due to neglecting lensing is seriously misestimated.
  Indeed, the first-order formalism that we use to estimate the shift in the best fits due to a systematic error is only valid to the extent that the shift is small compared to the errors (error contours are themselves meaningful only close enough to the fiducial cosmology), and also assuming that the systematic error does not affect the ellipse contours \cite{Kitching:2008eq}.
  Neither of these conditions is actually satisfied.

\subsection{MCMC with Planck priors}

A more realistic analysis makes use of prior knowledge of parameters from previous experiments. We therefore
repeat our MCMC analysis using Planck priors for all the cosmological parameters except the bias, which is not measured in Planck, and the neutrino mass. The latter is our most interesting parameter, and we want to test how strongly it is biased in an analysis which neglects lensing.

\begin{figure*}[bthp]
  \centering
  \includegraphics[scale=1.2]{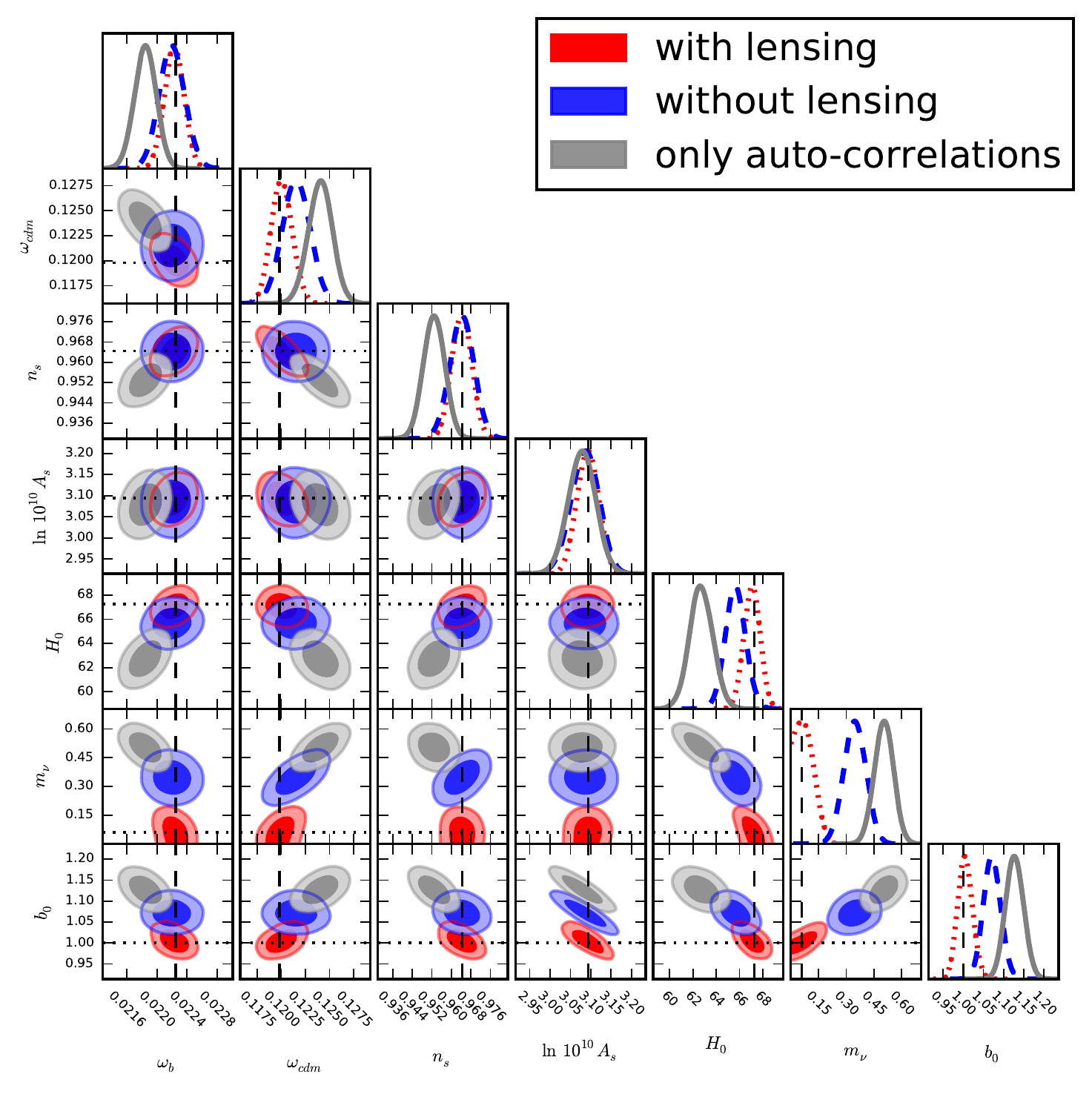}
  \caption{2D and 1D posteriors for the cosmological parameters inferred using Planck priors. We show the full analysis including lensing (red dotted), an analysis neglecting lensing (blue dashed), and considering only autocorrelations (gray solid).
    The 68\% and 95\% confidence intervals are shown.
    Intersections between vertical and horizontal lines denote the fiducial cosmology.
   See Table \ref{Table:2} for numerical values of the statistical quantities.
  }
  \label{fig:mcmc-cmb-prior}
\end{figure*}

Planck chains are publicly available through the \textit{Planck Legacy Archive}. In this paper, we use the chain for the extended model with a free neutrino mass based on the Planck TT, TE, EE + lowP likelihoods (Eq. (54c) in \cite{Ade:2015xua}). We compute the covariance matrix $\mathbf{C}$ for the cosmological parameters $\vec{x}=(\omega_b,\omega_{\mathrm{cdm}},n_s,A_s,H_0)$ and assume a Gaussian distribution for the prior. The $\chi^2$ relative to the fiducial model including the Planck prior is then the $\Delta \chi^2$ in Eq. \eqref{eq:chi2} plus
\begin{equation}
\Delta \chi^2_{\mathrm{prior}} = \sum_{i,j} (x_i - x^{\mathrm{fid}}_i)^2 C^{-1}_{ij} (x_j - x^{\mathrm{fid}}_j)^2,
\label{Eq:chi2-prior}
\end{equation}
where $\vec{x}^{\mathrm{fid}}$ denotes parameters of the fiducial model and $\mathbf{C^{-1}}$ is the inverse of the covariance matrix.
In this way, we marginalize the Planck prior over the neutrino mass and the optical depth, $\tau$, which are parameters that we want to leave free since we want to determine the first and our survey is not sensitive to the second. The results are shown in Table~\ref{Table:2} and Fig.~\ref{fig:mcmc-cmb-prior}.

Cosmological parameters in this case are clearly better determined than for the case without priors.
While the spectral index $n_s$ shows now a smaller relative shift, the neutrino masses and galaxy bias actually acquire larger shifts.
The incompatibility of the data and model pulls the Hubble parameter $H_0$ away from the fiducial value by over $4\sigma$ in spite of the Planck prior.
Hence, while the details of the analysis are important in determining the actual size of error bars and degeneracies in parameter space, a large bias $2\sigma$--$9\sigma$ in the neutrino masses is a feature that persists in all the analyses here performed.

\begin{table}[!t]
  \centering
  \begin{tabular}{@{}cccccc}
    \hline
    \multicolumn{6}{c}{i) Consistently including lensing: $\Delta \chi^2 = 0$} \\
    \hline
    Parameter & Mean & Best fit & $\sigma$ &\hspace{-0.52cm} shift: Mean & Best fit \\
    \hline
    $\omega_b$ & $0.02223$ & $0.02226 $ &$0.00013 $ &  \quad$0.2\sigma$ & $ <0.1\sigma$ \\
    $\omega_{cdm}$ & $0.1200 $ & $0.1196 $ & \quad$0.0011 $ &  \quad$0.2\sigma$ & $0.2\sigma$ \\
    $n_s$      & $0.9642 $ & $0.9651 $ & $0.0041 $ &  \quad$0.1\sigma$ & $ 0.1\sigma$ \\
    $\ln10^{10}A_s$ & $3.092 $ & $3.098$ & $0.026 $ &  \quad$0.1\sigma$ & $ 0.2\sigma$ \\
    $H_0\left(\frac{\text{km}}{\text{s}\cdot\text{Mpc}}\right)$      & $67.08$ & $67.25$ & $0.70$ &  \quad$0.3\sigma$ & $ <0.1\sigma$ \\
    $m_{\nu}$\,(eV)  & $0.08$ & $0.04$ & $0.05$ & \quad $ 0.4\sigma$ & $ 0.4\sigma$ \\
    $b_0$ & $1.005$ & $0.994$ & $0.018$ & $0.3\sigma$ & $0.3\sigma$ \\
  \end{tabular}
  \begin{tabular}{@{}cccccc}
    \hline
    \multicolumn{6}{c}{ii) Neglecting lensing: $\Delta \chi^2 = 2082$} \\
    \hline
    Parameter & Mean & Best fit & $\sigma$ & \hspace{-0.52cm} shift: Mean & Best fit \\
    \hline
    $\omega_b$ & $0.02220$ & $0.02219 $ & $0.00017 $ &  \quad$0.3\sigma$ & $0.4\sigma$ \\
    $\omega_{cdm}$ & $0.1215$ & $0.1214$ & $0.0014$ &  \quad$1.2\sigma$ & $1.1\sigma$ \\
    $n_s$      & $0.9643$ & $0.9640$ & $0.0049$ &  \quad$<0.1\sigma$ & $0.1\sigma$ \\
    $\ln10^{10}A_s$ & $ 3.085 $ & $3.090 $ & $ 0.034 $ &  \quad$0.3\sigma$ & $0.1\sigma$ \\
    $H_0\left(\frac{\text{km}}{\text{s}\cdot\text{Mpc}}\right)$      & $65.66$ & $65.64$ & $0.87$ &  \quad$1.8\sigma$ & $1.9\sigma$ \\
    $m_{\nu}$\,(eV)  & $0.35$ & $0.34$ & $0.06$ &  \quad$4.8\sigma$ & $4.7\sigma$ \\
    $b_0$ & $1.072$ & $1.070$ & $0.022$ & $3.3\sigma$ & $3.3\sigma$\\
  \end{tabular}
  \begin{tabular}{@{}cccccc}
    \hline
    \multicolumn{6}{c}{\parbox[t]{4.4cm}{iii) Neglecting lensing: \\ \hspace*{0.9cm} (only autocorrelations)} $\Delta \chi^2 = 230$} \\
    \hline
    Parameter & Mean & Best fit & $\sigma$ & \hspace{-0.52cm} shift: Mean & Best fit\\
    \hline
    $\omega_b$ & $0.02185 $ & $0.02181 $ & $0.00014 $ &  \quad$2.8\sigma$ & $3\sigma$ \\
    $\omega_{cdm}$ & $0.1240 $ & $0.1240 $ & $0.0013 $ &  \quad$3.4\sigma$ & $3.3\sigma$ \\
    $n_s$      & $0.9529 $ & $0.9536 $ & $0.0044 $ &  \quad$2.7\sigma$ & $2.5\sigma$ \\
    $\ln10^{10}A_s$ & $3.079 $ & $3.081 $ & $0.033 $ &  \quad$0.5 \sigma$ & $0.4\sigma$ \\
    $H_0\left[\frac{\text{km}}{\text{s}\cdot\text{Mpc}}\right]$      & $62.72 $ & $62.71$ & $1.01$ &  \quad$4.5 \sigma$ & $4.5\sigma$ \\
    $m_{\nu}$\,[eV]  & $0.50$ & $0.52$ & $0.05$ &  \quad$8.6\sigma$ & $8.8\sigma$ \\
    $b_0$ & $1.127$ & $1.127$ & $0.022$ & $5.7\sigma$ & $5.7\sigma$ \\
  \end{tabular}
 \caption{MCMC results with Planck priors. We show the mean and best-fit values, the standard deviation, and the amplitude of the shift of the mean and best fit with respect to the fiducial value in units of the standard deviation, $\si$, of the corresponding analysis. The large value of $\De\chi^2$ for case ii shows that cross-correlations cannot be fitted if lensing is neglected.
  }
  \label{Table:2}
\end{table}

\subsection{Significance of the lensing detection}

We can quantify the strength with which we detect the lensing signal in our setup with the help of Bayesian model probabilities, comparing the case with lensing to the case without lensing. To do this, we introduce formally an extended model $\MM_L$ with an additional ``lensing amplitude'' parameter $A_L$ that multiplies the lensing contribution in the model. For the ``with lensing'' model $\MM_1$, we then set $A_L = 1$, while the ``without lensing'' case $\MM_0$ corresponds to $A_L=0$. In this way, the two models are nested within the extended model, and we can use the Savage-Dickey density ratio (SDDR) method to derive model probabilities (see e.g., Ref. \cite{Trotta:2005ar} for an explanation of the SDDR and Sec. 3 of Ref. \cite{Dirian:2016puz} for a more detailed description of the same reasoning as that used here); with the SDDR, the Bayes factor $B$ between the case with fixed $A_L$ and the general case is given by the posterior for $A_L$ (marginalized over all other parameters) of the general model divided by prior, both taken at the nested point,
\begin{equation}
B_x \equiv \frac{P(D|\MM_x)}{P(D|\MM_L)} = \frac{P(A_L=x|D,\MM_L)}{P(A_L=x|\MM_L)} \, .
\end{equation}
Here, $P$ denotes probabilities, $D$ denotes the data and $x$ is either 0 or 1.
The Bayes factor between two models with given fixed values for $A_L$ is then simply the ratio of the Bayes factors relative to the extend model,
\begin{eqnarray}
B_{xy} &\equiv& \frac{P(D|\MM_x)}{P(D|\MM_y)} \nonumber \\
&=& \frac{P(D|\MM_x)}{P(D|\MM_L)} \frac{P(D|\MM_L)}{P(D|\MM_y)} = \frac{B_x}{B_y} \nonumber \\
 &=& \frac{P(A_L=x|D,\MM_L)}{P(A_L=y|D,\MM_L)} \, ,
\end{eqnarray}
where the last equality holds if $P(A_L=x|\MM_L)=P(A_L=y|\MM_L)$, e.g., for a uniform prior in $A_L$, which is what we will use. We see that the only information needed to determine $B_{xy}$ is the relative value of the posterior at $A_L=x$ and at $A_L=y$, and this is approximately given by the $\chi^2$ difference between these cases. As by construction $A_L=1$ (the case where we include lensing consistently) has $\Delta\chi^2=0$, we find simply that $\ln B_{01} \approx -\Delta\chi^2_{\mathrm{no~lensing}}/2$. We find thus that $\ln B_{01} \approx -1000$ when using auto- and cross-correlations and $\ln B_{01} \approx -90$ to $-115$ when only taking into account autocorrelations. Both Bayes factors are way out on the often-used Jeffreys scale \cite{Jeffreys:1939xee} where anything larger than 5 is considered as strong. In other words, lensing is detected in both cases with overwhelming evidence.

We can also translate the $\Delta\chi^2$ value into an order-of-magnitude estimate of ``the number of sigmas'' with which we detect the lensing signal in our setup. Assuming a Gaussian probability distribution function for $A_L$ so that $\Delta\chi^2 \approx (A_L-1)^2/\sigma[A_L]^2$, we find that $\sigma[A_L]$ needs to be 0.022 in order to explain the observed $\Delta\chi^2$ values of 2064 and 2082. This implies that the lensing is measured roughly at the $45\sigma$  level. Lensing is clearly a strong signal in the photo-$z$ type survey that we have considered here. As also discussed above, most of the lensing signal is contained in the off-diagonal spectra. The $\Delta\chi^2$ values of 180 and 230 when only looking at the autocorrelations correspond to about $13\sigma$ to $15\sigma$, roughly comparable to the strength of the lensing detection in the Planck temperature power spectrum \cite{Ade:2015xua}.

This also confirms the result of Ref. \cite{Montanari:2015rga}, which found that the lensing amplitude $A_L$ can be determined to an accuracy of the order of (1--2)\%  with a Euclid like photometric survey, with the constraints coming especially from the off-diagonal (interbin) correlations.

\section{Conclusions}

In this paper, we have shown that neglecting  lensing convergence leads to large shifts in the best-fit values of cosmological parameters for the data sets
available from future surveys. As in the CMB, where the lensing of the power spectra is detected at over $10\sigma$ \cite{Ade:2015xua},
it will become mandatory to include lensing also in the analysis of galaxy surveys.

In the case studied here, we have seen mainly an increase in the neutrino mass $m_\nu$ and a decrease in the spectral index $n_s$ when neglecting lensing. Also, the product $A_sb_0^2$ which determines the amplitude of fluctuations increases. This comes from the fact that the magnification bias for the Euclid specifications is relatively large~\cite{Montanari:2015rga} (see also Appendix~\ref{apa}), so that the  density-lensing correlation in bins with $z>1$ contributes with a positive sign. At smaller redshifts, which mainly measure correlations on smaller scales, this has to be corrected since there the total lensing term $\propto (5s-2)\kappa$ contributes negatively. This can be achieved by lowering $n_s$ and increasing the neutrino mass.

We note that
the specific shifts which we have obtained in our analysis depend  on the details of the survey. The main, generic result is that, in order to estimate cosmological parameters reliably with future galaxy surveys, we have to correctly include lensing with the measured magnification bias function, $s(z)$, defined by
$$ s(z)  \equiv \frac{\partial\log_{10} N(z,m<m_*)}{\partial m_*}\,,  $$
where $m_*$ is the limiting magnitude of the survey and $N(z,m)$ is the galaxy luminosity function of the survey at redshift $z$.

The fact that deep galaxy surveys are so sensitive to lensing, however, is not only a curse but also a blessing. It means that these surveys will allow us to determine a map of the lensing potential at different redshifts, i.e.\ perform ``lensing tomography'' with galaxy clustering. This will be a very interesting alternative to lensing tomography with shear measurements proposed, e.g.,\
in Ref. \cite{Heavens:2003jx}. Both techniques are challenging but they have different systematic errors and allow valuable cross-checks. So, clearly both paths should be pursued.

\begin{acknowledgments}
  It is a pleasure to thank Stefano Camera, Julien Lesgourgues,  and Roy Maartens for helpful comments and interesting  discussions.
  R. D. and M. K. acknowledge financial support from the Swiss National Science Foundation. W. C. is supported by the Departamento Administrativo de Ciencia, Tecnolog\'{i}a e Innovaci\'{o}n (Colombia).
  The calculations for this article were performed on the Baobab cluster of the University of Geneva.
     Plots were realized with {\sc GetDist} \footnote{\url{https://github.com/cmbant/getdist}.
  }
\end{acknowledgments}

\appendix

\section{The Euclid photometric survey}
\label{apa}

Angular power spectra, depending on two redshifts $z_i$ and $z_j$, can be written as integrals of transfer functions $\Delta_{\ell}^i(k)$ over wave numbers $k$:
\begin{equation} \label{eq:Cl}
C_{\ell}^{ij} = 4\pi \int d\ln k\; \mathcal{P}_{\mathcal{R}}(k) \Delta_{\ell}^{i}(k) \Delta_{\ell}^{i}(k) \;.
\end{equation}
Here, $\mathcal{P}_{\mathcal{R}}(k)=A_s k^{n_s-1}$ is the primordial power spectrum of curvature perturbations.
The transfer functions $\Delta_{\ell}^i(k)$ include an integral over a window function $W_i(z)$ describing the binning in redshift, multiplied by the number of galaxies per redshift interval $dN/dz$:
\begin{equation} \label{eq:Delta_l}
  \Delta_{\ell}^i(k) = \int dz\; \frac{dN}{dz} W_i(z) \Delta_{\ell}(z,k) \;.
\end{equation}
The main contributions to the transfer functions $\Delta_{\ell}(z,k)$ appearing in the integral of Eq.~(\ref{eq:Delta_l}) are given by the intrinsic galaxy density perturbation, redshift space distortions, and lensing effects:
\begin{eqnarray}
  \Delta_{\ell}(z,k) &=& b_G(z) \delta(z,k) j_\ell(kr(z))
  + \frac{k}{\cal H} V(z,k) \frac{d^2 j_\ell(kr(z))}{d(kr(z))^2}
  \nonumber \\
  &&+ \left( \frac{2-5s}{2}\right) \ell(\ell+1)
  \nonumber \\
  &&\times \int_0^{r(z)} d\tilde r\; \frac{r(z)-{\tilde r}}{r(z) {\tilde r}} \left[ {\Phi}(\tilde z,k) + {\Psi}(\tilde z,k) \right] j_\ell(k{\tilde r}) \;.
  \nonumber \\
\end{eqnarray}
We introduced the Fourier transforms of the density perturbations (in comoving gauge), of the metric perturbations $\Phi$, $\Psi$ and of the velocity potential, $v_i\equiv-\partial_i V$, in the Newtonian gauge\footnote{With initial conditions such that ${\cal R}(z_{\rm{in}},k)=1$.}.
The functions $j_\ell(kr(z))$ denote the spherical Bessel functions.
The integral along the line of sight describes the effects of lensing convergence which affects number counts by magnifying the sources, hence affecting their number density per steradian.
The factor $s(z)$ is called the magnification bias, and it depends on the luminosity function of the given galaxy population.
Note that for the special value $s=2/5$ lensing has no effect on number counts, while it has opposite sign for larger or smaller values, respectively.

\begin{figure}[t!]
\vspace{0.3cm}
\begin{center}
\includegraphics[width=.45\textwidth]{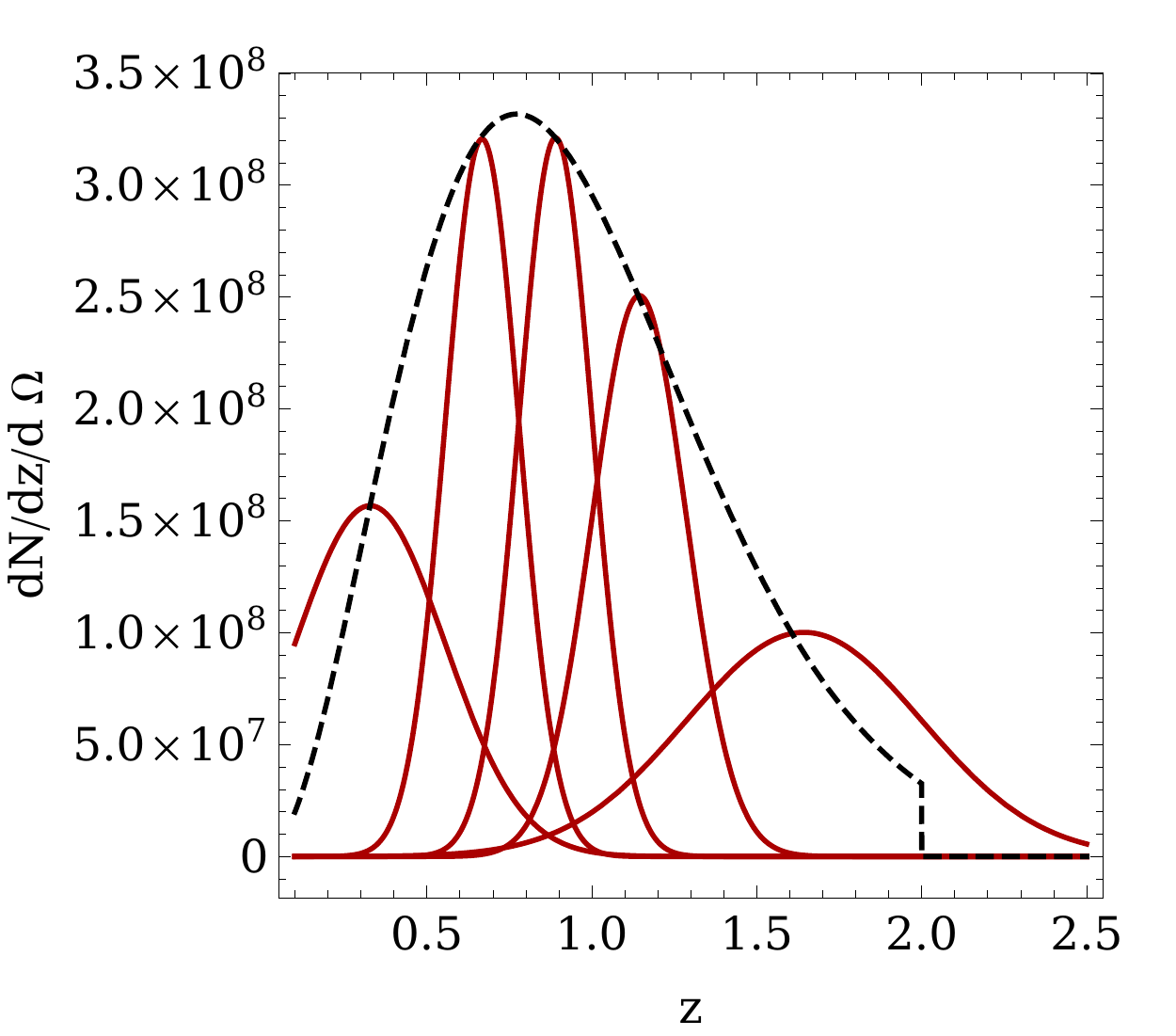}
\end{center}
\caption{Euclid photometric galaxy density distribution (black line) with a division into five bins containing the same number of galaxies.
}
\label{fig:dNdz}
\end{figure}

\begin{figure}[t!]
\begin{center}
\includegraphics[width=.45\textwidth]{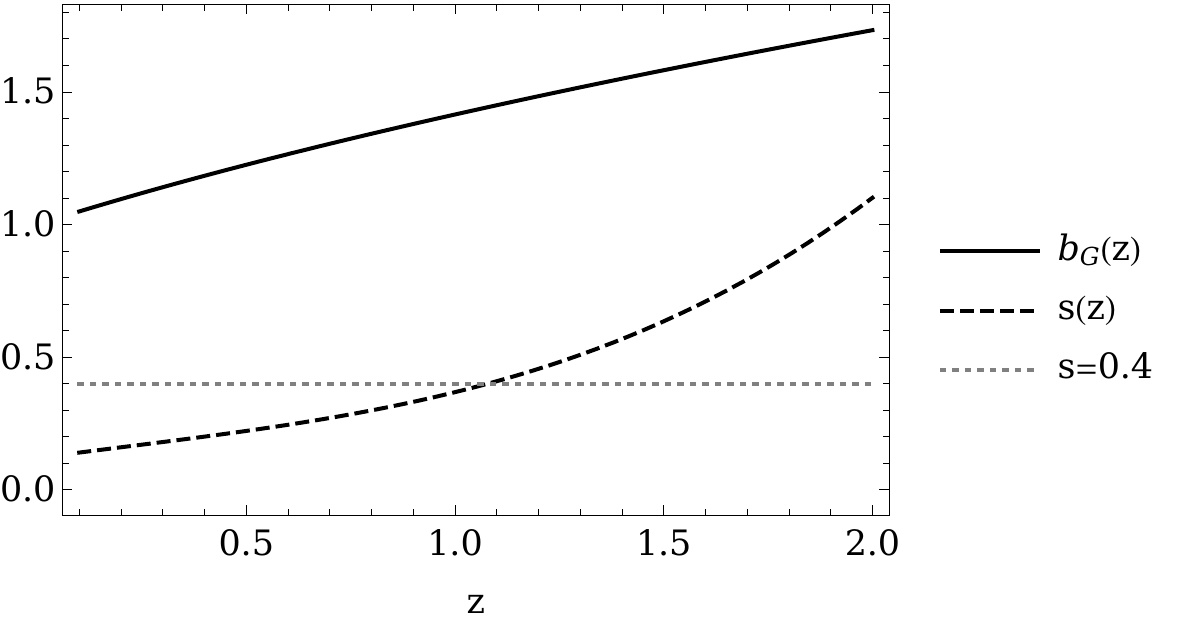}
\end{center}
\caption{Galaxy bias $b_G(z)$ and magnification bias $s(z)$ for Euclid. The
magnification bias is computed at the limiting magnitude $m_{\rm lim}=24.5$.
As a reference, we also plot the value $s=0.4$ at which the lensing contribution to number counts changes sign.
}
\label{fig:bs}
\end{figure}

Following Refs. \cite{Laureijs:2011gra,Amendola:2016saw}, we consider Euclid photometric specifications and approximate the number of galaxies per redshift and per steradian, the galaxy density, the covered sky fraction, the galaxy bias, and magnification bias as
\bea
&&\frac{dN}{dzd\Omega} = 3.5\times10^8 z^2 \exp\left[-\left( \frac{z}{z_0} \right)^{3/2}\right] \; \\
&&\quad \mbox{for} \quad 0<z<2.0\;, \nonumber \\
&&d=30\mbox{ arcmin}^{-2}\;,\\
&&f_{\rm sky}=0.364\;,\\
&&b_G(z)=b_0\sqrt{1+z}\;,\\
&&s(z)=s_0 + s_1 z + s_2 z^2 + s_3 z^3 \;, \label{eq:sz_euclid}
\eea
where $z_0=z_{\rm mean}/1.412$ and the median redshift is $z_{\rm mean}=0.9$.
We set $b_0=1$ in our fiducial model and then vary it in the MCMC chains.
The magnification bias is computed in Ref.~\cite{Montanari:2015rga}, and the coefficients are $s_0=0.1194$, $s_1=0.2122$, $s_2=-0.0671$, and $s_3=0.1031$.
Figure~\ref{fig:dNdz} shows the division into five Gaussian bins containing the same number of galaxies.
For numerical convenience, we set the lower redshift bound to $z=0.1$; this affects  our results by a negligible amount.
Figure~\ref{fig:bs} shows the redshift dependence of galaxy and magnification bias. We assume constant galaxy bias and magnification bias within each bin, the values being determined by the mean redshift of the bin.

\vspace{1cm}
\section{Basic expressions for the Fisher analysis\label{app:fisher}}

The Fisher approach used in the literature~\cite{Namikawa:2011yr,Duncan:2013haa,Camera:2014sba,Raccanelli:2015vla} and applied in Sec. ~\ref{s:fisher} for comparison with the results from our MCMC forecasts is based on the Fisher information matrix given by
\begin{equation}
F_{\alpha\beta} =
\sum_{\ell} \sum_{(ij)(pq)} \frac{\partial C_\ell^{ij}}{\partial \theta_\alpha}
\frac{\partial C_\ell^{pq}}{\partial
\theta_\beta} {{\rm Cov}_{C_{\ell \, \rm[(ij), (pq)]}}^{-1}} \, ,
\label{eq:Fisher}
\end{equation}
where $\theta_a$ denotes a given cosmological parameter.
We compute the derivatives with a five-point stencil \cite{Montanari:2015rga}, and the derivative step for each parameter is set with an iterative procedure to be of the same size as the 1-$\sigma$ levels obtained when fixing the other parameters $\sigma_{\theta_{\alpha}}=1/\sqrt{F_{\alpha\alpha}}$.
We verified that the final results do not depend significantly on the particular step values.
We sum up to $\ell=400$, while the second sum is over the matrix indices $(ij)$ with $i \leq j$ and $(pq)$ with $p \leq q$ which run from 1 to the total number of bins when all bin auto- and cross-correlations are taken into account.
Using the same notation as in Eq.~(\ref{eq:Cl_th_obs}), the covariance matrix is
\begin{equation}
\label{eq:err-clgt}
{\rm Cov}_{C_{\ell \, \rm[(ij), (pq)]}} = \frac{C^{\rm A, (ip)}_{\ell} C^{\rm A, (jq)}_{\ell} + C^{\rm A, (iq)}_{\ell} C^{\rm A, (jp)}_{\ell}}{(2\ell+1)f_{\rm sky}}.
\end{equation}
If only autocorrelations are taken into account, the covariance must be first reduced to the relevant components and subsequently inverted.
We estimate the shift in the best-fit values due to the wrong model assumption $\widetilde{C}_{\ell}$ by defining the systematic error as $\De C_{\ell}=C_{\ell}-\widetilde{C}_{\ell}$ \cite{Knox:1998fp,Heavens:2007ka,Kitching:2008eq,Camera:2014sba},
\begin{equation}
\label{eq:shift}
\Delta_{\theta_{\alpha}}=\sum_{\beta} \left[\left(\widetilde{F}\right)^{-1}\right]_{\alpha\beta} B_{\beta} \;,
\end{equation}
where we defined
\begin{equation}
B_{\beta} = \sum_{(ij)(pq)} \sum_{\ell} \De C_{\ell}^{ ij} \frac{\partial \widetilde{C}_{\ell}^{pq}}{\partial \theta_{\beta}} {\rm Cov}_{\widetilde{C}_{\ell \, \rm[(ij), (pq)]}}^{-1} \;.
\end{equation}
A tilde always denotes the quantity computed according to the wrong model $\widetilde{C}_{\ell}$.
This expression assumes that the systematic error does not affect the covariance, and it is only valid if the shifts are small compared to the variances $\Delta^2_{\theta_{\alpha}}/\sigma^2_{\theta_{\alpha}}<1$.
As mentioned in the text, neither of these hypothesis is satisfied in our case.
Furthermore, note that Eq.~(\ref{eq:Fisher}) can only be used to estimate error contours by assuming that the underlying universe is described either by $C_{\ell}$ or by $\tilde{C}_{\ell}$ and does not give information about error contours obtained when  fitting the wrong model $\tilde{C}_{\ell}$ to data consistent with the full $C_{\ell}$ spectra.

\bibliographystyle{utcaps}
\bibliography{biblio}

\end{document}